\pdfoutput=1
\PassOptionsToPackage{table,xcdraw}{xcolor}

\documentclass[11pt]{article}
\usepackage{acl}
\usepackage{times}
\usepackage{amsmath}
\usepackage{pgffor}
\usepackage{enumitem}
\usepackage{graphicx}
\usepackage{xspace}
\usepackage{comment}
\usepackage{booktabs}
\usepackage{multirow}
\usepackage{makecell}
\usepackage[table,xcdraw]{xcolor}
\usepackage{listings}
\usepackage{pifont}
\usepackage{tcolorbox}
\usepackage{subcaption}
\usepackage{balance}

\newlength{\ColorBoxDepthReference}
\newlength{\ColorBoxHeightReference}
\newlength{\Width}%
\newcommand{\MyColorBox}[2][red]%
{%
	\settowidth{\Width}{#2}%
	\colorbox{#1}%
	{%
		\raisebox{-\ColorBoxDepthReference}%
		{%
			\parbox[b][\ColorBoxHeightReference+\ColorBoxDepthReference][c]{\Width}{\centering#2}%
		}%
	}%
}
\newcommand{\best}[1]{\fboxsep1.5pt\colorbox{gray!50}{#1}}
\newcommand{\second}[1]{\fboxsep1.5pt\colorbox{gray!25}{#1}}

\newcommand{\dmd}[1]{\textcolor{red}{\{\textbf{DD:} #1\}}}

\newcommand{\ie}{\textit{i.e.,}\xspace}
\newcommand{\eg}{\textit{e.g.,}\xspace}

\newcommand{\bench}{\textsc{JitVul}\xspace}

\title{Benchmarking LLMs and LLM-based Agents\\ in Practical Vulnerability Detection for Code Repositories}

\author{
	Alperen Yildiz$^1$, Sin G. Teo$^2$, Yiling Lou$^3$, Yebo Feng$^4$, Chong Wang$^4$\thanks{Chong Wang is the corresponding author.}, Dinil Mon Divakaran$^2$\\
	$^1$National University of Singapore, Singapore\\
	$^2$Institute for Infocomm Research, A*STAR, Singapore\\
	$^3$Fudan University, China\\
	$^4$Nanyang Technological University, Singapore
}

\begin{document}

\maketitle
\begin{abstract}
Large Language Models (LLMs) have shown promise in software vulnerability detection, particularly on function-level benchmarks like Devign and BigVul. However, real-world detection requires interprocedural analysis, as vulnerabilities often emerge through multi-hop function calls rather than isolated functions. While repository-level benchmarks like ReposVul and VulEval introduce interprocedural context, they remain computationally expensive, lack pairwise evaluation of vulnerability fixes, and explore limited context retrieval, limiting their practicality.

We introduce \bench, a JIT vulnerability detection benchmark linking each function to its vulnerability-introducing and fixing commits. Built from 879 CVEs spanning 91 vulnerability types, \bench enables comprehensive evaluation of detection capabilities. Our results show that ReAct Agents, leveraging thought-action-observation and interprocedural context, perform better than LLMs in distinguishing vulnerable from benign code. While prompting strategies like Chain-of-Thought help LLMs, ReAct Agents require further refinement. Both methods show inconsistencies, either misidentifying vulnerabilities or over-analyzing security guards, indicating significant room for improvement.
\end{abstract}
\section{Introduction}
Given the success of large language models (LLMs) across various application domains, researchers have begun exploring their effectiveness in software vulnerability detection. On well-known vulnerability detection benchmarks such as Devign~\cite{zhou2019devign} and BigVul~\cite{fan2020ac}, LLMs—particularly those fine-tuned on code—have shown promising results, suggesting their potential for real-world applications.

However, a significant gap exists between these widely used benchmarks and the requirements for real-world vulnerability detection in code repositories~\cite{wang2024reposvul,wen2024vuleval}. These benchmarks primarily focus on function-level vulnerability detection, where a single function is input to a detector for label prediction without considering the broader repository context. In contrast, real-world vulnerabilities—such as null pointer dereference (NPD)—often arise within multi-hop function call chains, and not in isolated functions. Detecting such vulnerabilities requires tracing interprocedural call relationships and understanding the relevant code elements like branch conditions~\cite{risse2024top}.

To address these limitations, recent studies have shifted towards repository-level detection scenarios, enabling more realistic benchmarking of LLMs for vulnerability detection. ReposVul~\cite{wang2024reposvul} and VulEval~\cite{wen2024vuleval} are two benchmarks that enhance the detection process by extracting callers and callees for a target function from the code repository, providing interprocedural context. These callers (functions that call the target function) and callees (functions called by the target function) are selectively fed into LLMs to assess the vulnerability of the target function. Findings from these benchmarks show that while LLMs benefit from the additional interprocedural context, they still exhibit low effectiveness in real-world scenarios, particularly when fine-tuning is not applied.

\begin{table}[]
\centering
\caption{Comparison of \bench with existing benchmarks for repository-level vulnerability detection.}\label{tab:bench-comp}
\renewcommand{\arraystretch}{1.1}
\setlength{\tabcolsep}{2pt}
\small
\vspace{-2mm}
\begin{tabular}{lccccc}
\toprule
\textbf{Benchmark}    & \textbf{\# CVEs} & \textbf{\# CWEs} & \textbf{Pairwise} & \textbf{Agents Eval} \\ \hline
ReposVul              &     6,134        &     236          & \ding{55}         & \ding{55}            \\
VulEval               &     4,196        &     5            & \ding{55}         & \ding{55}            \\
\bench (ours)         &     879          &     91           & \ding{51}         & \ding{51}            \\ 
\bottomrule
\end{tabular}
\vspace{-3mm}
\end{table}

Although existing works have provided valuable insights into the effectiveness of LLMs for repository-level vulnerability detection, several key limitations remain in achieving more comprehensive benchmarking.
First, in existing repository-level vulnerability detection approaches, all functions within a code repository are treated as target functions for vulnerability detection. This approach becomes computationally expensive and impractical, particularly for large repositories like the Linux kernel. 
Second, the benchmarks do not effectively assess the capability of LLMs in distinguishing between vulnerable functions and those where the vulnerability has been patched. As highlighted by a recent study~\cite{risse2024uncovering}, this is a critical limitation of machine learning-based vulnerability detection methods.
Finally, the integration of interprocedural context (\ie callers and callees) has mostly been limited to retrieval-based strategies, leaving many potential approaches underexplored. LLM-based agentic methods, such as ReAct~\cite{yao2022react}, offer the potential for on-demand, iterative acquisition of interprocedural context, enabling more adaptive analysis.

To bridge the gap, we target the task of \textit{\textbf{just-in-time (JIT) vulnerability detection}}~\cite{lomio2022just}, a more practical approach for identifying vulnerabilities in code repositories. Unlike prior methods that analyze all functions in a repository, JIT vulnerability detection is triggered only for functions modified in a commit, with interprocedural context provided. Inspired by prior research~\cite{risse2024uncovering}, we construct a pairwise benchmark called \bench for JIT vulnerability detection, where each target function is linked to both a vulnerability-introducing commit and a vulnerability-fixing commit. To achieve this, we first select 879 Common Vulnerabilities and Exposures (CVE) entries, each representing a unique vulnerability, from PrimeVul, a high-quality function-level detection dataset~\cite{ding2024vulnerability}. We then extract target functions from the vulnerability-fixing commits explicitly referenced in the CVE entries, obtaining both their vulnerable and patched versions. Finally, we analyze the commit history of each vulnerable function to identify the corresponding vulnerability-introducing commit. The resulting \bench comprises 1,758 paired commits spanning 91 Common Weakness Enumerations (CWEs).

We implement LLMs and ReAct Agents with various prompting strategies and foundation models to assess their effectiveness in JIT vulnerability detection. Our evaluation on \bench uncovers several key findings. A higher \textit{F1} score doesn’t always reflect a method's ability to capture vulnerability characteristics, highlighting the need for pairwise evaluation. ReAct Agents, using their thought-action-observation framework and interprocedural context, better differentiate between vulnerable and benign versions. While strategies like CoT and few-shot examples boost LLMs' performance, ReAct Agents need more tailored designs. Both LLMs and ReAct Agents often exhibit inconsistent analysis patterns between the vulnerable and benign versions, indicating a lack of robustness in vulnerability analysis. These findings highlight key areas for future research in vulnerability detection: (i) developing more comprehensive evaluation guidelines for benchmarking LLMs and LLM-based agents in JIT vulnerability detection, (ii) exploring advanced prompting strategies for improving LLM-based agents in vulnerability detection, and (iii) designing robust reasoning models tailored to vulnerability analysis that capture the true essence of vulnerabilities, rather than relying on speculation.

This paper makes the following contributions:
\begin{itemize}[leftmargin=15pt]
    \item We introduce \bench, a benchmark for just-in-time (JIT) vulnerability detection in code repositories, consisting of 1,758 pairwise commits spanning 91 vulnerability types.
    \item We implement ReAct agents with various prompting strategies and foundation models and evaluate their effectiveness in leveraging interprocedural context for JIT detection.
    \item Our experimental results provide valuable insights into the application of LLMs and LLM-based agents for real-world vulnerability detection. We explore the necessity of pairwise evaluation, the advantages and disadvantages of LLMs and ReAct agents, and the impact of prompting designs and foundation models.
    \item We release all code and data at \href{https://anonymous.4open.science/r/JitVul-C6C7/}{this repository}.

\end{itemize}



\section{Related Work}

\subsection{Vulnerability Detection Benchmarks}
Several benchmarks have been proposed for function-level vulnerability detection. BigVul~\cite{fan2020ac} collects C/C++ vulnerabilities from the CVE database, filtering out entries without public Git repositories, and labels functions as vulnerable or non-vulnerable based on commit fixes. MegaVul~\cite{ni_megavul_2024} improves upon existing benchmarks by using code parsing tools for accurate function extraction and de-duplicating functions referenced by multiple CVEs. DiverseVul~\cite{chen_diversevul_2023} ensures data quality by filtering vulnerability-introducing commits with specific keywords and deduplicating function bodies with hash functions. PrimeVul~\cite{ding2024vulnerability} addresses data quality challenges by proposing filtering rules to handle noise labels and duplicated functions.

For repository-level vulnerability detection, ReposVul~\cite{wang2024reposvul} addresses issues with tangled and outdated patches, using trace-based filtering to ensure data quality and integrating repository-level features to provide richer context for detection. VulEval~\cite{wen2024vuleval} provides a framework that collects high-quality data from sources like Mend.io Vulnerability Database~\cite{mendio} and National Vulnerability Database~\cite{nvd}, including contextual information like caller-callee relationships. 

\subsection{LLM-based Vulnerability Detection}
Recent studies have explored the use of Large Language Models (LLMs) for vulnerability detection, highlighting their ability to enhance both the identification and explanation of software vulnerabilities. 
LLM4SA~\cite{wen2024automatically} integrates language models with SAST tools, leveraging LLMs to inspect static analysis warnings and significantly reduce false positives. LLM4Vuln~\cite{sun_llm4vuln_2024} enhances LLM execution by incorporating more context through a retrieval-augmented generation (RAG) pipeline and static analysis. LSAST~\cite{keltek2024lsast} further explores context augmentation, employing multiple RAG pipelines to compare the effectiveness of retrieval augmentation using static analysis outputs, vulnerability reports, and code abstraction. Similarly, Vul-RAG~\cite{du2024vul} constructs a vector database of vulnerability reports alongside a language model engine. \cite{zhou2024comparison} introduce a voting mechanism that combines SAST tools and LLMs for vulnerability detection.

\subsection{LLMs and LLM-based Agents}
Various methods have been explored to enhance LLM performance, with prompt augmentation being a key approach that enriches prompts to improve the model’s reasoning process. Chain-of-thought (CoT) prompting~\cite{wei2022chain} is one of the most widely used techniques, where instructions like ``Let's think step by step'' guide the model to break problems into sub-problems. Few-shot prompting~\cite{brown2020language} is another common method, providing example traces to enable in-context learning without modifying model weights. Both CoT and few-shot prompting are frequently employed in LLM-based vulnerability detection~\cite{zhou2024comparison, wen2024automatically}.

Agentic architectures are among the most promising state-of-the-art technologies but remain underexplored in vulnerability detection~\cite{zhou2024large}. \cite{yao2022react} introduce Reasoning and Acting (ReAct) agents, which generate reasoning and action traces in an interleaved manner to interact with their environment and analyze resulting observations. This iterative process continues until the agent determines a final answer. Reflexion agents~\cite{shinn_reflexion_2023} are similar to ReAct agents however they focus on self-reflection and dynamic memory updates alongside with reinforcement learning. Self-refine agents~\cite{madaan_self-refine_nodate} are another type of agents where the same language model is instructed to provide feedback based on the output. There have also been several multi-agent systems, such as Alpha-Codium~\cite{ridnik_code_nodate}, where the agents are represented as nodes on graphs.
\section{\bench: Just-in-Time Vulnerability Detection for Code Repositories}
In this section, we discuss the requirements of benchmarking LLMs and LLM-based agents for repository-level vulnerability detection and formulate the task of just-in-time (JIT) vulnerability detection. We also present a benchmark for JIT detection, derived from real-world vulnerabilities.

\subsection{Problem Statement}  
Benchmarking vulnerability detection in real-world code repositories requires considering three key practicality requirements:  

\begin{itemize}[leftmargin=15pt]  
\item \textbf{Interprocedural Context.} Many vulnerabilities originate from interprocedural interactions, even though their manifestation and required fixes often occur within individual functions~\cite{wang2024reposvul}. For instance, a null pointer dereference (NPD) vulnerability may arise when a pointer initialized as null in one function is improperly dereferenced in another function along the execution path. Identifying such vulnerabilities necessitates analyzing interprocedural dependencies, as examining functions in isolation is insufficient. 

\item \textbf{Scalability.} A straightforward application of learning-based methods to vulnerability detection for code repositories entails scanning each function individually and predicting a binary label. However, in the context of LLMs and LLM-based agents, this approach becomes computationally infeasible for large-scale repositories due to the high processing costs and resource constraints associated with analyzing extensive codebases. A more practical strategy is to focus on a limited set of candidate functions. \textit{Just-in-time} vulnerability detection~\cite{lomio2022just} exemplify this by prioritizing functions that have been newly introduced or modified in commits.

\item \textbf{Pairwise Comparison.} Traditional evaluation methods for machine learning-based vulnerability detection present models with labeled vulnerable code alongside other functions in the repository. However, recent findings~\cite{risse2024uncovering} suggest that models struggle to distinguish between vulnerable code and its patched, benign version, indicating an over-reliance on superficial patterns rather than meaningful vulnerability indicators. This highlights the necessity of pairwise benchmarking to ensure reliable evaluation of vulnerability detection methods.  
\end{itemize}  

Although recent works address some of these requirements, to the best of our knowledge, no study fully satisfies all three. For example, PrimeVul~\cite{ding2024vulnerability} is a high-quality dataset that provides pairwise evaluation for LLMs, but it focuses on function-level detection and does not account for interprocedural context. On the other hand, VulEval~\cite{wen2024vuleval} is a repository-level detection benchmark, but it lacks pairwise evaluation and considers only callers and callees modified in the same commit as the relevant interprocedural context when extracting dependencies—an assumption that is not always valid.


\subsection{Task Definition}
We define the task of \textit{just-in-time vulnerability detection} as follows. Given a code repository $\mathcal{R}$ and a target function $f$ modified in a commit, the task is formulated as:  
$$  
\textsc{JitDetect}: (\mathcal{R}, f) \to \{\texttt{vul}, \texttt{ben}\},  
$$  
where $f$ is classified as either vulnerable (\texttt{vul}) or benign (\texttt{ben}) based on the (interprocedural) context within $\mathcal{R}$. Building on this, we propose a pairwise benchmark to evaluate LLMs and LLM-based agents.

\subsection{Benchmark Construction}
\begin{figure}
    \centering
    \includegraphics[width=1\columnwidth]{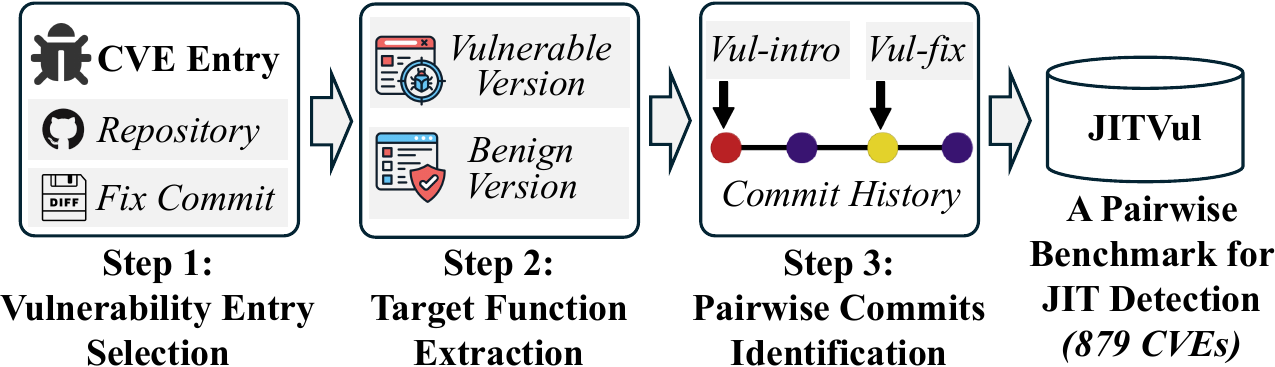} 
    \caption{Construction process of \bench.}
    \label{fig:bench-construction}
    \vspace{-5mm}
\end{figure}

We construct a benchmark called \bench for practical Just-In-Time (JIT) vulnerability detection for code repositories, building on the function-level detection dataset PrimeVul~\cite{ding2024vulnerability}. As presented in Figure~\ref{fig:bench-construction}, the construction process involves three key steps: Vulnerability Entry Selection, Target Function Extraction, and Pairwise Commit Identification.

\textbf{Step 1: Vulnerability Entry Selection.}
We begin by selecting CVE entries that meet three key criteria: (i) the selected CVEs should cover a broad range of CWEs, representing different categories of vulnerabilities, (ii) each CVE should have an associated GitHub repository with a complete commit history, and (iii) each CVE should correspond to a commit that fixes the vulnerability. To satisfy these criteria, we leverage the PrimeVul dataset as our foundation. PrimeVul ensures high data quality by including only CVEs that were fixed in a single commit modifying a single function. From PrimeVul, we randomly select 879 CVE entries, ensuring coverage across 91 CWEs, with a maximum of ten CVEs per CWE.

\textbf{Step 2: Target Function Extraction.}
For each selected CVE, we retrieve the target function and the commit that fixes the vulnerability directly from the PrimeVul dataset. The versions before and after the fix correspond to the vulnerable and benign versions of the target function, denoted as $f_{vul}$ and $f_{ben}$, respectively.

\textbf{Step 3: Pairwise Commits Identification.}
Unlike function-level vulnerability detection studies like PrimeVul, our JIT detection requires identifying the commits that trigger vulnerability detection for both $f_{vul}$ and $f_{ben}$, and obtaining the corresponding repository versions to provide necessary context like callers and callees. To achieve this, we extract the two commits responsible for introducing and fixing the vulnerability, referred to as the \textit{vul-intro} and \textit{vul-fix} commits, respectively.

The \textit{vul-fix} commit can be directly obtained from the data retrieved in the previous step, and the repository version corresponding to it is denoted as $\mathcal{R}_{fix}$. Identifying the precise \textit{vul-intro} commit is more challenging, as pinpointing the exact code change that introduces the vulnerability requires considering the complex interactions between functions. Following the methodology from prior JIT detection work~\cite{lomio2022just}, we trace the change history of the target function to approximate the \textit{vul-intro} commit. Specifically, we traverse the commit history backward, examining each commit until we identify the commit where the target function was last modified to become the $f_{vul}$ version. This commit is then designated as the \textit{vul-intro} commit, and the corresponding repository is denoted as $\mathcal{R}_{intro}$.

In the \textit{vul-intro} and \textit{vul-fix} commits, the target function is modified into the $f_{vul}$ and $f_{ben}$ versions, respectively, thereby triggering the JIT detection process.

\textbf{Resulting Benchmark.}
After completing the above steps, \bench includes 1,758 pairwise data samples, with 879 labeled as vulnerable and 879 as benign. Each sample consists of a specific code repository version, a target function, and a ground-truth label (\ie \texttt{vul} or \texttt{ben}). These samples are derived from 879 CVE entries, spanning 91 CWEs. The average number of code files in the repositories is 2,955.94, and the average number of lines of code in the target functions is 696.40.

\section{Experimental Setup}
\subsection{Studied Methods}
We investigate three categories of detection methods: Plain LLM, Dependency-Augmented (Dep-Aug) LLM, and ReAct Agent. We choose the ReAct agent over alternatives because its thought-action-observation workflow aligns well with the need for on-demand interprocedural analysis in JIT vulnerability detection.

\begin{itemize}[leftmargin=15pt]  
    \item \textbf{Plain LLM} employs a single LLM with a prompt for vulnerability detection, resembling a function-level detection approach. The LLM is given only the target function to determine whether it is vulnerable. The detailed prompts can be found in Appendix Section~\ref{sec:llm-prompts}.
    
    \item \textbf{Dep-Aug LLM} extends the Plain LLM by incorporating Top-5 similar callers and callees of the target function into the prompt through a lexical retrieval approach, such as Jaccard Similarity. This method, proposed and evaluated in VulEval~\cite{wen2024vuleval}, is reproduced in our work based on the original paper. We use it as a baseline that integrates interprocedural context into LLMs in a deterministic manner.
    
    \item \textbf{ReAct Agent} performs an iterative thought-action-observation process as illustrated in Figure~\ref{fig:react} in Appendix, which is equipped with three tools for on-demand interprocedural context acquisition: (i) \textbf{get\_callers} returns the function names and line numbers of callers for the input function; (ii) \textbf{get\_callees} returns the function names and line numbers of callees for the input function; (iii) \textbf{get\_definition} retrieves the complete function definition based on the input function name and line number. Using these three tools, our ReAct agent for JIT detection follows the workflow outlined below. In each iteration, the agent first reasons based on the current context and the observations from the previous iteration. It then decides whether to call the tools for interprocedural context or to stop the iteration and make a prediction. After observing the tool outputs, the agent proceeds to the next iteration.
    
\end{itemize}

For each category, besides the \textit{vanilla} version with a basic prompt, we also design three variants based on different prompting strategies: \textit{chain-of-thought (CoT)}, \textit{few-shot examples (FS)}, and \textit{combination of both CoT and FS}.
\begin{itemize}[leftmargin=15pt]  
    \item \textbf{CoT}: We use a basic CoT approach by adding the instruction ``Solve this problem step by step...'' to prompt the LLM and ReAct agent to break down the reasoning tasks. We avoid more complex CoT instructions, as summarizing reasoning patterns for vulnerabilities in advance is challenging, particularly given the wide variety of vulnerability types (CWEs).
    \item \textbf{FS}: We provide several pairwise examples, each consisting of a vulnerable code snippet with a detailed explanation of the vulnerability and its patched benign version with an explanation of the applied safeguard. These few-shot examples are expected to offer context for the LLM and ReAct agent to distinguish between vulnerable and benign code, helping them focus on the actual vulnerability features.
\end{itemize}

For each variant, we further employ two different foundation models: GPT-4o-mini and GPT-4o.

\subsection{Metrics}
For each \textit{vul-intro} and \textit{vul-fix} commit pair, we apply detection methods to the corresponding repository ($\mathcal{R}_{intro}$ or $\mathcal{R}_{fix}$) and target function ($f_{vul}$ or $f_{ben}$), then compare predictions with ground truth. In addition to the commonly used \textbf{\textit{F1}} score, we also assess effectiveness using \textit{pairwise accuracy} (\textbf{\textit{pAcc}}), inspired by PrimeVul~\cite{ding2024vulnerability}. This metric reflects the proportion of pairs where both functions are correctly labeled, \ie $\textsc{JitDetect}(\mathcal{R}_{intro}, f_{vul}) = \texttt{vul}$ and $\textsc{JitDetect}(\mathcal{R}_{fix}, f_{ben}) = \texttt{ben}$.

$$F1 = \frac{2 \times TP}{2\times TP + FP + FN},$$
$$pAcc = \frac{\textit{\# of Correctly Labeled Pairs}}{\textit{\# of Total Pairs}},$$
\\where TP is the number of true positives, FN is the number of false negatives, and
FP is the number of false positives. 


\subsection{Implementation}
\textbf{Few-shot Example Creation.}
To support few-shot variants, we manually create ten example pairs for both the LLM and the agent. Each pair consists of a vulnerable example and a benign example, sourced from the 2024 CWE Top 25 Most Dangerous Software Weaknesses list\footnote{https://cwe.mitre.org/top25/index.html} on the CWE website. For each weakness in the Top 25 list, we review its corresponding web page to identify a relevant C/C++ code snippet. These snippets typically serve as illustrative examples of how the vulnerability manifests, often accompanied by detailed explanations. We use the original example as the vulnerable code, making minor edits to the explanation for normalization (\eg merging two paragraphs before and after the code into a single cohesive text). Subsequently, we manually modify the code snippet to address the vulnerability, drafting an explanation of why the modified code is benign, referencing the original explanation of the vulnerable code. In this manner, we create ten pairs of vulnerable and benign examples, each accompanied by its respective explanation. Appendix Figure~\ref{fig:few-shot} provides an illustrative example.

\textbf{Caller and Callee Extraction.}
Given a code repository and a function, we extract its callers and callees using CFlow~\cite{cflow}, a widely used tool for on-demand call graph construction. For each caller or callee identified by CFlow, we then use CTags~\cite{ctags} to extract its complete function body from its definition in the code repository. These are implemented as Python functions using LangChain's tool decorator for integration into the ReAct workflow.

\textbf{LLMs and Agents.}
We use GPT-4o-mini and GPT-4o with the temperature set to 0. LangChain-0.3.14 is used for pipeline construction.

\section{Results and Analyses}
Table~\ref{tab:main-results} presents the results of the studied detection methods on \bench.

\begin{table}[t]
    \caption{Results of studied methods on \bench. The \best{best} and \second{second-best} results are highlighted.}\label{tab:main-results}
    \centering
    \renewcommand{\arraystretch}{1.1}
    \setlength{\tabcolsep}{5pt}
    \small
    \begin{tabular}{llrrlrr}
    \toprule
    \multirow{2}{*}{\textbf{Method}} &  & \multicolumn{2}{c}{\textbf{GPT-4o-mini}}          &  & \multicolumn{2}{c}{\textbf{GPT-4o}}                \\ 
    \cline{3-4} \cline{6-7}
                                     &  & \textbf{\textit{F1}~~}  & \textbf{\textit{pAcc}} &  & \textbf{\textit{F1}~~}    & \textbf{\textit{pAcc}}  \\
    \midrule
    \textbf{Plain LLM}               &  &                         &                          &  &                         &                         \\        
    - vanilla                        &  & 56.00                   & 3.36                     &  & \best{65.96}            & 1.02                    \\ 
    - w/ CoT                         &  & \second{65.10}          & 3.36                     &  & 62.22                   & 15.02                   \\
    - w/ FS                          &  & 48.74                   & 7.56                     &  & 62.77                   & 4.44                    \\
    - w/ CoT+FS                      &  & 64.65                   & 11.76                    &  & \second{64.44}          & 17.63                    \\
    \midrule
    \textbf{Dep-Aug LLM}             &  &                         &                          &  &                         &                          \\ 
    - vanilla                        &  & 52.68                   & 2.05                     &  & 63.30                   & 1.03                    \\
    - w/ CoT                         &  & \best{66.05}            & 4.86                     &  & 62.60                   & 18.66                    \\
    - w/ FS                          &  & 48.23                   & 7.27                     &  & 62.03                   & 2.39                    \\
    - w/ CoT+FS                      &  & 65.01                   & 4.68                     &  & 61.12                   & 18.79                    \\    
    \midrule
    \textbf{ReAct Agent}             &  &                         &                          &  &                         &                          \\ 
    - vanilla                        &  & 56.63                   & 12.61                    &  & 57.77                   & 17.63                    \\
    - w/ CoT                         &  & 56.93                   & \second{16.81}           &  & 58.07                   & \best{19.13}             \\
    - w/ FS                          &  & 56.81                   & \best{20.17}             &  & 56.42                   & \second{18.91}           \\
    - w/ CoT+FS                      &  & 51.06                   & 14.29                    &  & 52.61                   & 18.89                    \\    
    \bottomrule
    \end{tabular}
\end{table}

\subsection{Detection Method Comparison}
We compare the three categories of detection methods based on both \textit{F1} and \textit{pAcc} scores.

\textbf{Results.} The results indicate that ReAct Agents achieve higher \textit{pAcc} scores than other LLM-based methods across all prompting strategies, with improvements ranging from 0.1\% to 16.61\%, with the largest gain occurring with GPT-4o and the vanilla prompting. However, Plain LLMs and Dep-Aug LLMs generally achieve higher \textit{F1} scores than ReAct under most settings, with improvements of 4.15\%-13.95\%, except when using GPT-4o-mini with the CoT and CoT+FS prompting strategies. Additionally, Dep-Aug LLMs do not show consistent improvements over Plain LLMs and even exhibit performance degradation with certain prompting strategies.

\textbf{Findings.} Two findings emerge from results. 

\textit{A higher \textit{F1} score does not necessarily indicate a detection method's superior ability to capture vulnerability characteristics.} Vulnerability detection methods exhibit an inconsistent relationship between \textit{pAcc} (pairwise accuracy) and \textit{F1} (isolated metric). LLM-based methods predict significantly more \texttt{vul} labels—exceeding 90\% in certain settings—compared to ReAct Agents, leading to higher recall. However, precision remains similar across methods, around 50\%, as observed on our \textit{label-balanced} benchmark.. These factors explain the higher \textit{F1} scores of LLM-based methods on \bench. This also highlights \textit{F1}'s sensitivity to the data distribution, emphasizing the need for pairwise evaluation to better capture core vulnerability characteristics~\cite{risse2024uncovering}.

\textit{The thought-action-observation framework of ReAct Agents, combined with their effective use of interprocedural context, enhances their ability to capture vulnerability characteristics.} ReAct Agents conduct in-depth, fine-grained analysis by iteratively and adaptively retrieving additional context, such as callers and callees, rather than relying on superficial analysis or speculation. This allows them to differentiate between code versions before and after vulnerability fix, leading to consistent improvements in \textit{pAcc}. In contrast, while Dep-Aug LLMs incorporate interprocedural context, they rely on mechanical retrieval based on similarity metrics (\eg Jaccard Similarity), feeding retrieved Top-5 callers and callees all at once, which may introduce noise. This could explain why Dep-Aug LLMs sometimes show a lower \textit{pAcc} than Plain LLMs. In comparison, ReAct agents demonstrate average improvements in \textit{pAcc} of 9.46\% over Dep-Aug LLMs with GPT-4o-mini and 8.42\% with GPT-4o, across various prompting strategies. To demonstrate the adaptive use of interprocedural context of ReAct Agent, we present the distribution of tool invocations for ReAct with GPT-4o and vanilla prompting in Appendix Figure~\ref{fig:distribution}. It shows that ReAct Agent dynamically invokes the tools one to three times to retrieve the necessary callers or callees for most cases, in contrast to Dep-Aug LLM, which feeds a fixed number of callers and callees.

\subsection{Prompting Strategy Comparison}
We also compare the three prompting strategies and perform a detailed analysis.

\textbf{Results.}
When applying different prompting strategies, such as CoT and FS, detection methods show varying degrees of improvement in \textit{pAcc} scores, ranging from 1.26\% to 17.76\%. However, these prompting strategies do not consistently lead to \textit{F1} improvements in the pairwise evaluation of JIT detection. 

\textbf{Findings.} We find two key findings.

\textit{Popular prompting strategies like CoT and FS examples can enhance LLMs' performance in pairwise JIT evaluation.} While these strategies sometimes reduce \textit{F1} scores, the pairwise metric \textit{pAcc} shows that LLMs can significantly benefit from CoT instructions (even something as simple as ``Solve this problem step by step...'') and pairwise FS examples. This improvement is often overlooked when focusing solely on \textit{F1} and should be considered when designing methods.

\textit{ReAct Agents require further design improvements when using prompting strategies.} The current prompts are straightforward and align better with the inference process of LLMs, meaning the improvements from these strategies for ReAct Agents are relatively smaller than for LLM-based methods. For instance, the FS examples consist of singleton code snippets that do not require interprocedural analysis, which limits the benefit for ReAct Agents that rely on interprocedural context. This suggests a research opportunity in developing agent-oriented prompting strategies specifically for vulnerability detection.

\subsection{Foundation Model Comparison}
To evaluate the effectiveness of different foundation models across various detection methods, we conduct a comparative analysis. Additionally, we include the additional open-source Llama3.1-8B for further comparison, with the results provided in Table~\ref{tab:llama-results} in the Appendix.

\textbf{Results.} GPT-4o outperforms GPT-4o-mini on average, while Llama-3.1-8B frequently fails to complete the analysis process and defaults to the \texttt{ben} label when integrated into ReAct Agents.

\textbf{Findings.} Two key findings are identified.

\textit{Different foundation models are sensitive to different prompting strategies.} The results show that GPT-4o-mini and GPT-4o exhibit distinct improvement patterns with CoT and FS examples. This highlights the need to customize prompting strategies based on the selected foundation models. Moreover, larger models do not always outperform smaller models, emphasizing the importance of carefully designing methods that consider the characteristics of specific foundation models.

\textit{The execution of ReAct Agents depends on the instruction-following capability of foundation models.} Inspection of the outputs reveals that the failures of Llama-3.1-8B in ReAct Agents are due to the model's frequent inability to follow output format requirements. This prevents the agents from linking outputs and inputs across components, thus failing to perform the thought-action-observation iterative framework. This is a known issue with some foundation models, which struggle to follow instructions effectively~\cite{verma2024brittlefoundationsreactprompting}, reducing their effectiveness when used in agentic architectures.

\subsection{Pairwise Comparison}
The failures in pairwise evaluation can be categorized into three types: \textbf{pairwise vulnerable}, where both versions are labeled as vulnerable; \textbf{pairwise benign}, where both versions are labeled as benign; and \textbf{pairwise reversed}, where both versions are mislabeled. We conduct a more detailed pairwise comparison based on these types.

\textbf{Results.} The most prevalent pairwise inaccuracy is \textit{pairwise vulnerable}, ranging from approximately 40\% to 95\% for LLMs and from around 35\% to 50\% for ReAct Agents (except those with Llama-3.1-8B). Typically, the occurrence of \textit{pairwise reversed} increases as \textit{pAcc} improves.

\textbf{Findings.} We identify the following insights.

\textit{ReAct Agents are more effective at distinguishing between pairwise target functions.}
ReAct Agents demonstrate a stronger ability to differentiate between a vulnerable function and its patched benign version. This is because they leverage the thought-action-observation framework to iteratively and adaptively retrieve additional context, such as callers and callees, which allows them to capture differences between the two versions. A related case study can be found in Appendix Section~\ref{sec:case-1}.

\textit{LLMs and ReAct Agents sometimes fail to identify the causes in the vulnerable version, while tending to over-analyze the benign version after the vulnerability fix.}
LLMs and ReAct Agents often struggle to pinpoint the root causes (\eg insufficient input sanitization) of vulnerabilities in the vulnerable version. After the vulnerability is fixed, however, the detection methods tend to over-analyze the patched guards (\eg newly added sanitization statements) in the benign version, speculating about results and often misidentifying non-issues. This discrepancy arises because LLMs rely on broad, general patterns without the accurate reasoning capability needed for complex contexts. This highlights the need for more fine-grained analysis capabilities, such as reliable constraint solving, which should be integrated with LLMs or enhanced through program analysis techniques. A detailed case study can be found in Appendix Section~\ref{sec:case-2}.

\textit{LLMs and ReAct Agents often exhibit inconsistent analysis patterns when analyzing pairwise target functions.}
When analyzing pairwise target functions—vulnerable and benign versions—LLMs display significant inconsistencies in their evaluation patterns. For some pairs, they may provide thorough analysis for the vulnerable version but neglect important details in the benign version, or vice versa. In other cases, they may exhibit contrasting reasoning or focus on irrelevant aspects, leading to discrepancies in how they interpret the two versions. A clear example of this inconsistency is the significant difference in the average number of tool invocations by the ReAct Agent for the vulnerable and benign versions (\eg 5.85 vs. 1.79 when using GPT-4o-mini with vanilla prompting). These inconsistencies highlight the challenges LLMs face in handling the complexities of code analysis, where the differences between a vulnerable and benign version can be subtle and require more nuanced evaluation. A detailed case study can be found in Appendix Section~\ref{sec:case-1}.

\section{Conclusion}
In this work, we introduced \bench, a benchmark for just-in-time (JIT) vulnerability detection that enables a comprehensive, pairwise evaluation of LLMs and LLM-based agents. Our results show that ReAct Agents, leveraging thought-action-observation and interprocedural context, demonstrate better reasoning but require further refinement, particularly in utilizing advanced prompting strategies. Additionally, LLMs and ReAct agents often misinterpret flaws by either overlooking critical issues or over-analyzing benign fixes. These findings highlight the need for improving agentic architectures, prompting techniques, dynamic interprocedural analysis, and robust reasoning models tailored to vulnerability analysis to enhance automated vulnerability detection.
\section{Limitations}
As the early work benchmarking LLM-based agents for JIT vulnerability detection in code repositories, we acknowledge several limitations. First, the construction of \bench may not perfectly trace the \textit{vul-intro} commit due to the complexity of code evolution and function interactions. To mitigate this, we followed existing methodologies~\cite{lomio2022just} and manually inspected selected instances to ensure reliability. Second, while we emphasize pairwise evaluation for JIT detection, the label-balanced dataset may introduce bias in F1 score comparisons. In the future, we plan to incorporate more benign commits to improve the evaluation in F1 metric. Third, some important statistics, such as the ratio of interprocedural vulnerabilities, are missing. Although we attempted manual annotation, it is labor-intensive and difficult to scale. We plan to leverage commercial annotation platforms to achieve the annotation and provide more fine-grained evaluations in future work.

\bibliography{ref}
\bibstyle{acl_natbib}
\balance

\clearpage
\onecolumn
\appendix
\section{Studied Methods}

\subsection{Prompt Templates in Plain LLM}\label{sec:llm-prompts}
\begin{figure}[h]
    \centering
    \includegraphics[width=0.9\textwidth]{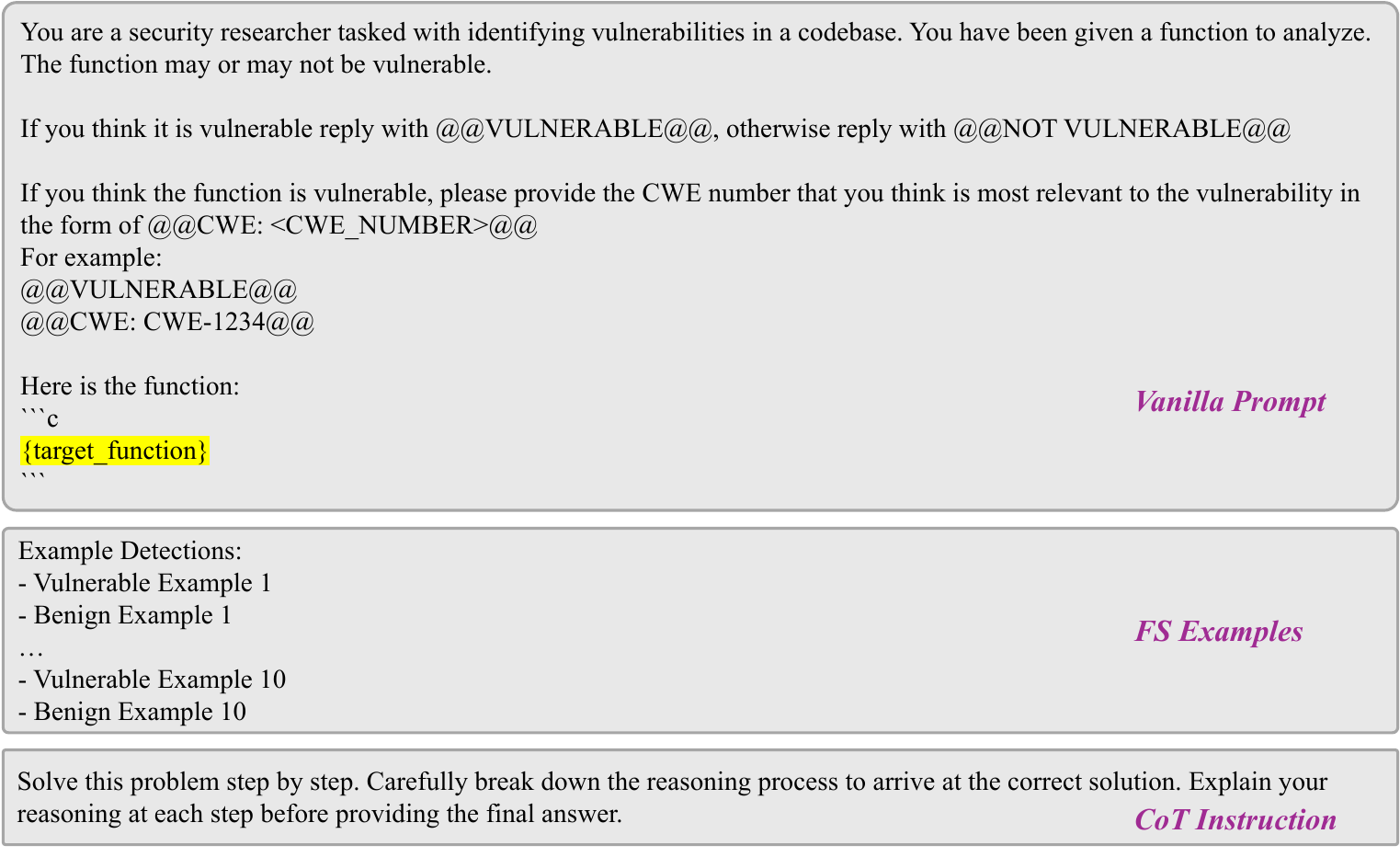}
    \caption{Prompt templates used with Plain LLM.}\label{fig:llm-prompts}
\end{figure}

Figure~\ref{fig:llm-prompts} illustrates the various prompting strategies used with Plain LLM. The \textit{Vanilla Prompt} serves as the base prompt included in all variants, while \textit{FS Examples} and \textit{CoT Instruction} are selectively applied according to the specific strategies. In the template, ``\{target\_function\}'' acts as a placeholder for the target function to be detected.

\clearpage

\subsection{ReAct Agent}
\begin{figure}[h]
    \centering
    \includegraphics[width=0.75\linewidth]{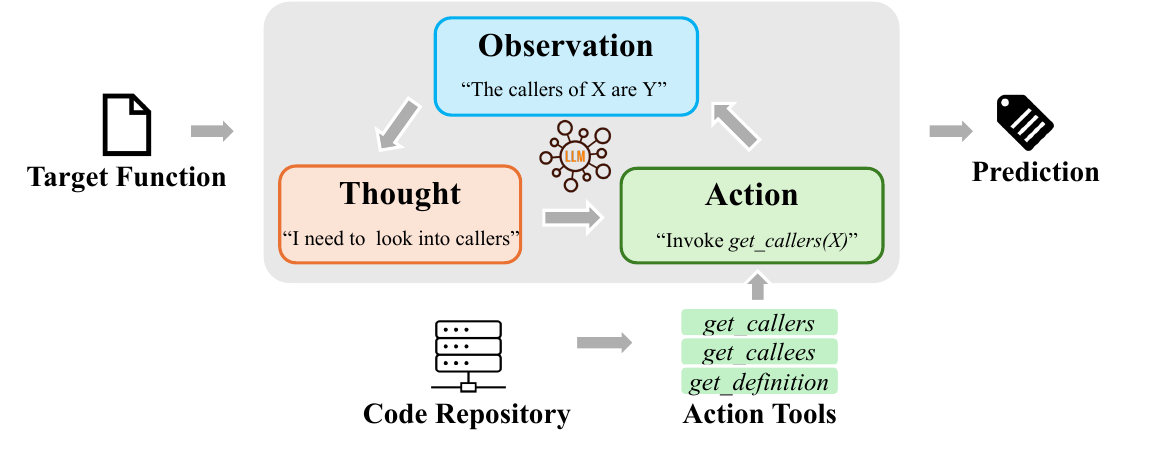}
    \caption{Workflow of ReAct Agent for JIT vulnerability detection.}
    \label{fig:react}
\end{figure}

\begin{figure}[h]
    \centering
    \includegraphics[width=0.7\textwidth]{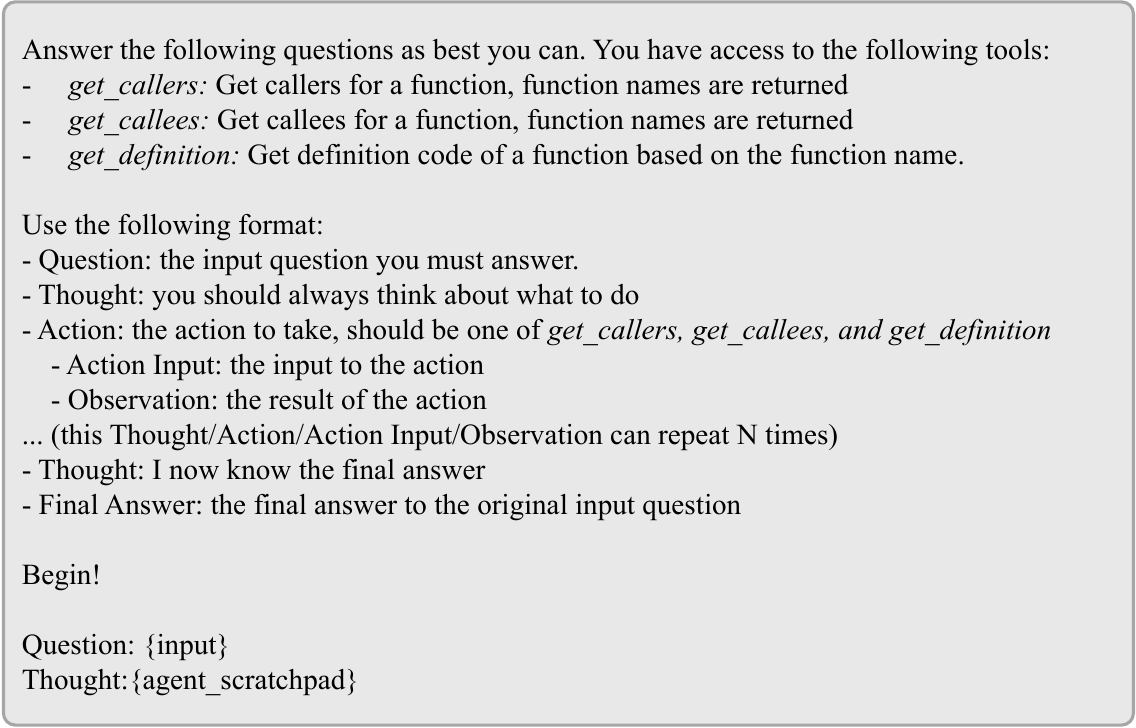}
    \caption{Prompt template used with ReAct Agent.}\label{fig:react-prompt}
\end{figure}

Figure~\ref{fig:react} illustrates the overall workflow of the ReAct Agent for JIT vulnerability detection. Figure~\ref{fig:react-prompt} displays the prompt used with the ReAct Agent, implemented with the default LangChain framework. The ``\{agent\_scratchpad\}'' is a one-time execution memory that holds tool descriptions, along with previous observations and reasoning traces. The ``\{input\}'' variable is used for the user prompt and can be further enhanced using prompt augmentation techniques.

\clearpage

\subsection{Few-shot Example}
\begin{figure}[h]
    \centering
    \begin{subfigure}[b]{0.45\textwidth}
        \centering
        \includegraphics[width=\textwidth]{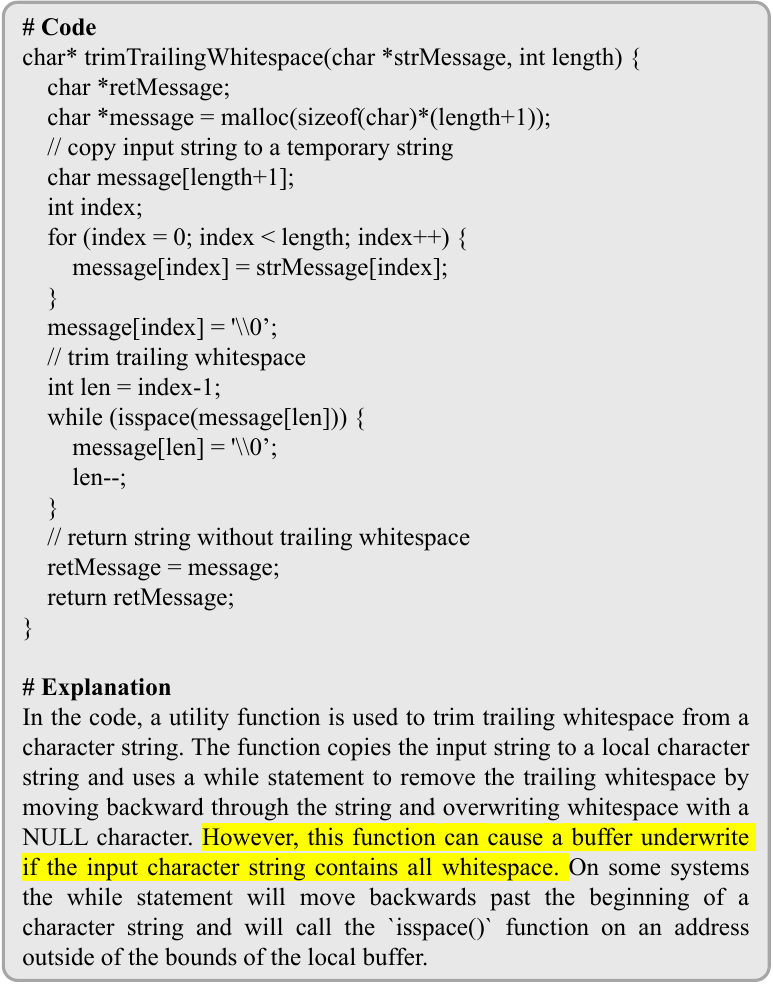}  
        \caption{Vulnerable Version}
        \label{fig:sub1}
    \end{subfigure}
    \hspace{0.5cm}  
    \begin{subfigure}[b]{0.45\textwidth}
        \centering
        \includegraphics[width=\textwidth]{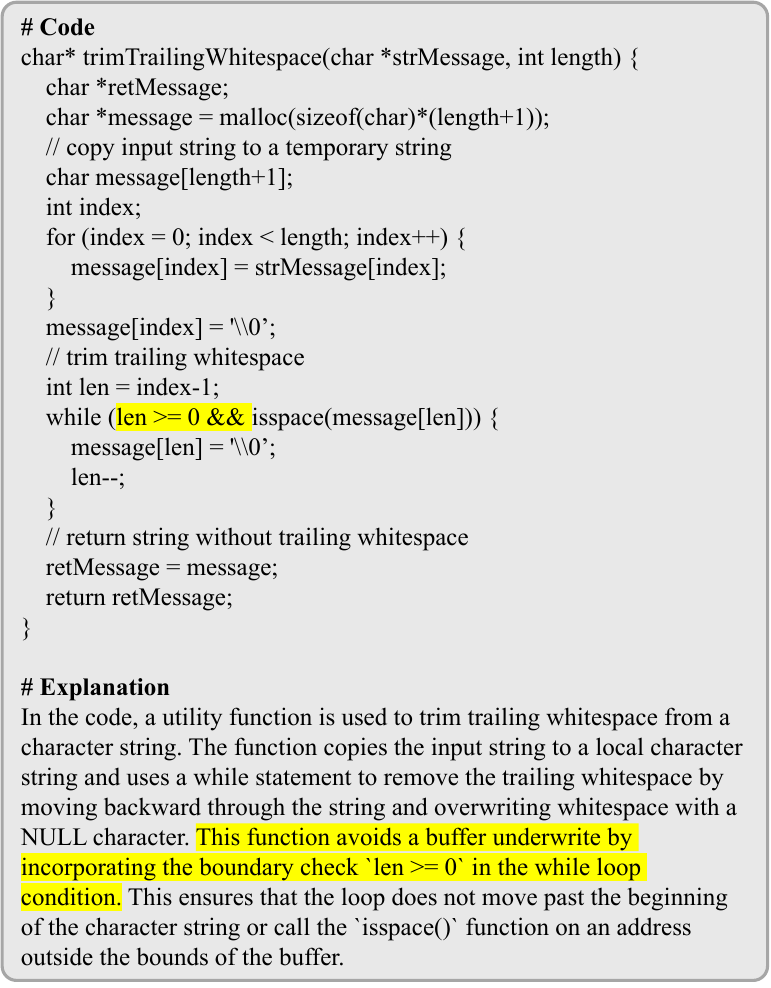}  
        \caption{Benign Version}
        \label{fig:sub2}
    \end{subfigure}
    \caption{A FS example of ``CWE-787: Out-of-bounds Write'', including both vulnerable version and benign version.}\label{fig:few-shot}
\end{figure}

Figure~\ref{fig:few-shot} presents a few-shot example based on the webpage for ``CWE-787: Out-of-bounds Write''. It highlights the key differences between the vulnerable and benign versions.

\clearpage

\section{Tool Invocation Distribution}
\begin{figure}[h]
    \centering
    \includegraphics[width=0.6\textwidth]{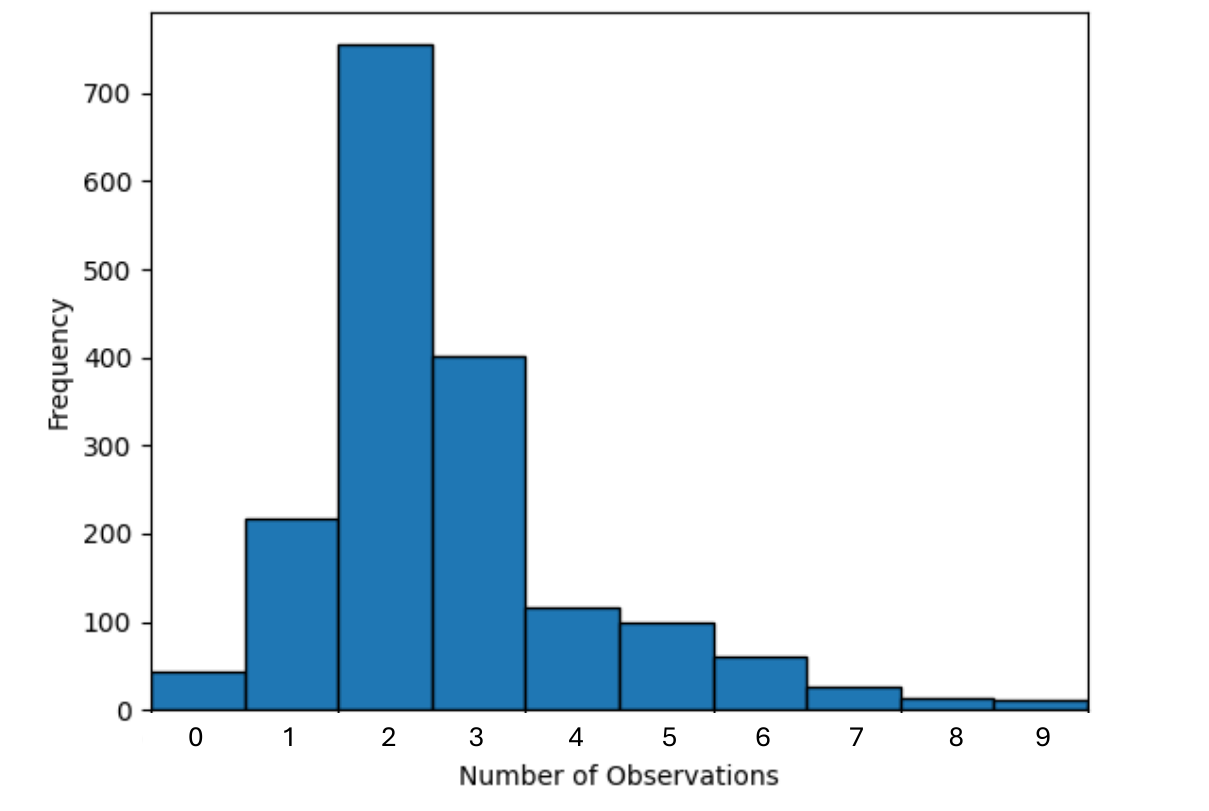}
    \caption{Distribution of tool invocations for ReAct Agent with GPT-4o and vanilla prompting.}\label{fig:distribution}
\end{figure}

Figure~\ref{fig:distribution} illustrates the distribution of tool invocations for ReAct Agent with GPT-4o and vanilla prompting. The data shows that, in most cases, ReAct Agent invokes the tools one to three times to retrieve the necessary callers or callees.

\section{Llama-3.1 Results}
Table~\ref{tab:llama-results} presents the results of Llama3.1-8B on \bench. ReAct Agents using Llama3.1-8B show significantly lower performance, with the execution process often failing due to formatting and parsing issues. As a result, the agents frequently default to the \texttt{ben} label.

\begin{table}[h]
    \caption{Results of Studied Detection Methods on \bench with Llama3.1-8B}\label{tab:llama-results}
    \centering
    \renewcommand{\arraystretch}{1.1}
    \setlength{\tabcolsep}{3pt}
    \small
    \begin{tabular}{llrrr}
    \toprule
    \textbf{Method}                  &  & \textbf{\textit{F1}~~}  & \textbf{\textit{pAcc}}     \\ 
    \midrule
    \textbf{Plain LLM}               &  &                         &                            \\        
    - vanilla                        &  & 58.05                   & 0.84                      \\
    - w/ CoT                         &  & 49.79                   & 10.92                      \\
    - w/ FS                          &  & 54.48                   & 1.68                      \\
    - w/ CoT+FS                      &  & 29.55                   & 14.29                      \\
    \midrule
    \textbf{Dep-Aug LLM}             &  &                         &                            \\ 
    - vanilla                        &  & 40.48                   & 15.17                      \\
    - w/ CoT                         &  & 21.18                   & 8.39                      \\
    - w/ FS                          &  & 27.37                   & 10.88                      \\
    - w/ CoT+FS                      &  & 16.46                   & 7.42                      \\    
    \midrule
    \textbf{ReAct Agent}             &  &                         &                            \\ 
    - vanilla                        &  & 9.09                    & 4.20                      \\
    - w/ CoT                         &  & 14.67                   & 3.36                      \\
    - w/ FS                          &  & 3.28                    & 0.84                      \\
    - w/ CoT+FS                      &  & 3.28                    & 1.68                      \\    
    \bottomrule
    \end{tabular}
\end{table}

\clearpage

\section{Case Study}~\label{appendix:case-study}
We provide several examples to illustrate the inputs and outputs of the detection methods for a better understanding of the analysis.

\subsection{CVE-2019-15164}\label{sec:case-1}
\begin{figure}[h]
    \centering
    \includegraphics[width=1\textwidth]{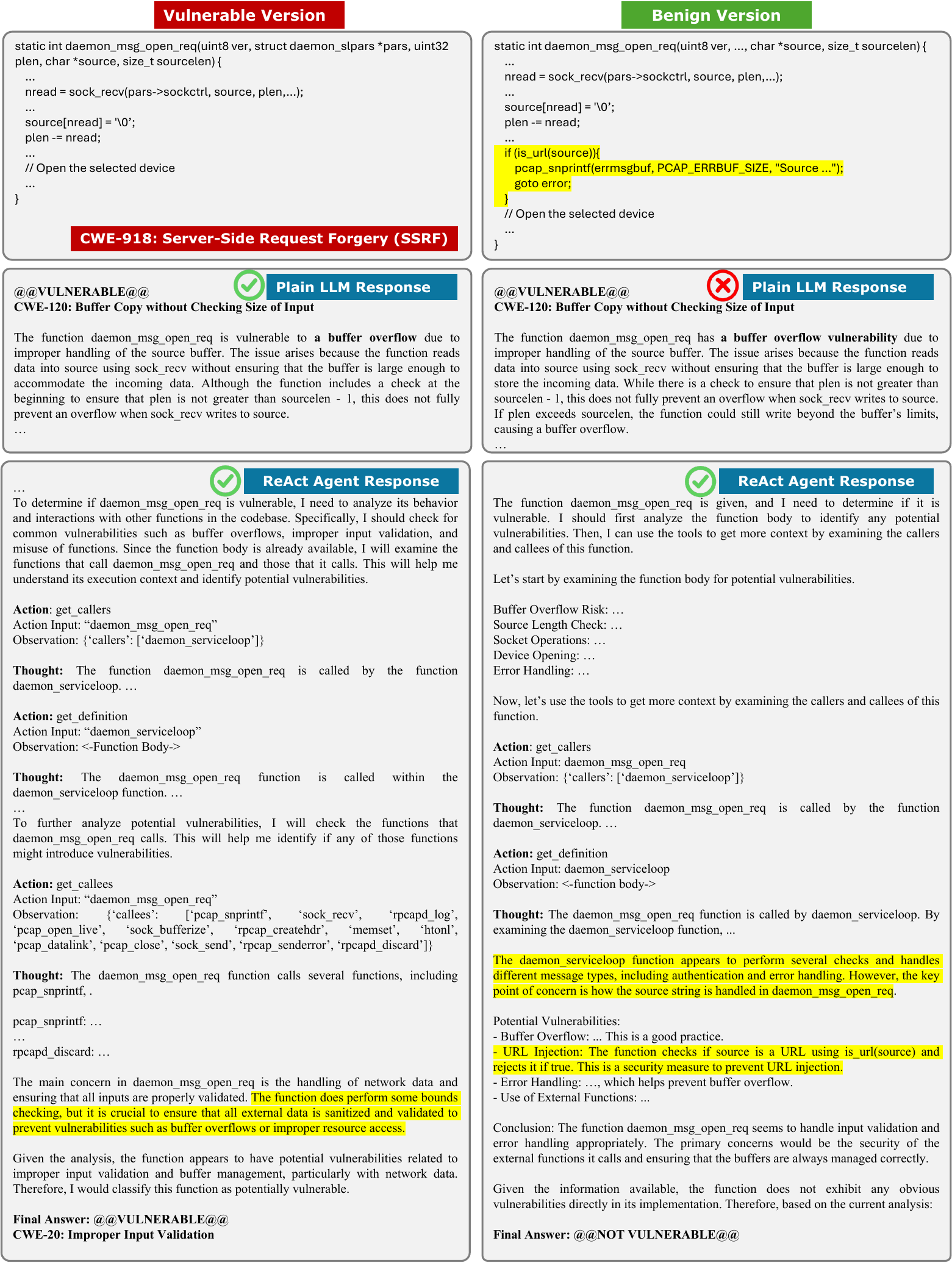}
    \caption{Case study for CVE-2019-15164.}\label{fig:case-1}
\end{figure}

Figure~\ref{fig:case-1} illustrates the case study derived from CVE-2019-15164 (details at https://nvd.nist.gov/vuln/detail/CVE-2019-15164), with the left side showing the vulnerable code and the detection methods' responses, and the right side depicting the benign version and its corresponding responses. The vulnerable version of the function \texttt{daemon\_msg\_open\_req} is susceptible to a ``CWE-918: Server-Side Request Forgery (SSRF)'' vulnerability due to the lack of validation for \texttt{source} before opening the device, which is read from the network socket. The benign version addresses this vulnerability by adding an if-condition to validate whether \texttt{source} is a valid URL, as highlighted in the figure.

\textbf{Label Predictions.} When using Plain LLM with GPT-4o and vanilla prompting, the analyses of both the vulnerable and benign versions focus on buffer operations and misclassify the benign as vulnerable. In contrast, when using the ReAct Agent, the predictions for both versions are correct. The agent is able to retrieve and analyze additional context, such as understanding its caller function \texttt{daemon\_serviceloop} and surrounding function bodies. This contextual information enables the agent to better comprehend how the \texttt{daemon\_msg\_open\_req} function is used within the broader codebase and recognize the risk introduced by the unvalidated URL input. Key points in the analysis process are highlighted to show the improved detection capability provided by the ReAct Agent.

\textbf{CWE Predictions.} However, upon examining the specific vulnerability categories predicted by Plain LLM and ReAct Agent, some fine-grained issues emerge. Plain LLM incorrectly predicts ``CWE-120: Buffer Copy without Checking Size of Input'' for both the vulnerable and benign versions, which is entirely inaccurate. On the other hand, ReAct Agent predicts ``CWE-20: Improper Input Validation'' for the vulnerable version. While this is not the correct classification, it is somewhat related to the ground-truth vulnerability of Server-Side Request Forgery (SSRF). The SSRF vulnerability arises from the improper validation of the \texttt{source} parameter before opening device, which the ReAct Agent's prediction partially captures, indicating a closer alignment to the actual issue.

\textbf{Analysis Patterns.} When delving into the detailed analysis processes, we observe that the ReAct Agent does not maintain consistent analysis patterns across both versions. For the vulnerable version, the agent focuses on buffer operation and input validation, while for the benign version, it conducts a more comprehensive check. However, in this case, the analysis patterns should be more similar, suggesting that the LLM behind the ReAct Agent lacks sufficient robustness to capture the actual vulnerability characteristics. This indicates a deficiency in its reasoning capabilities for accurate vulnerability reasoning.

\subsection{CVE-2019-3877}\label{sec:case-2}
\begin{figure}[h]
    \centering
    \includegraphics[width=1\textwidth]{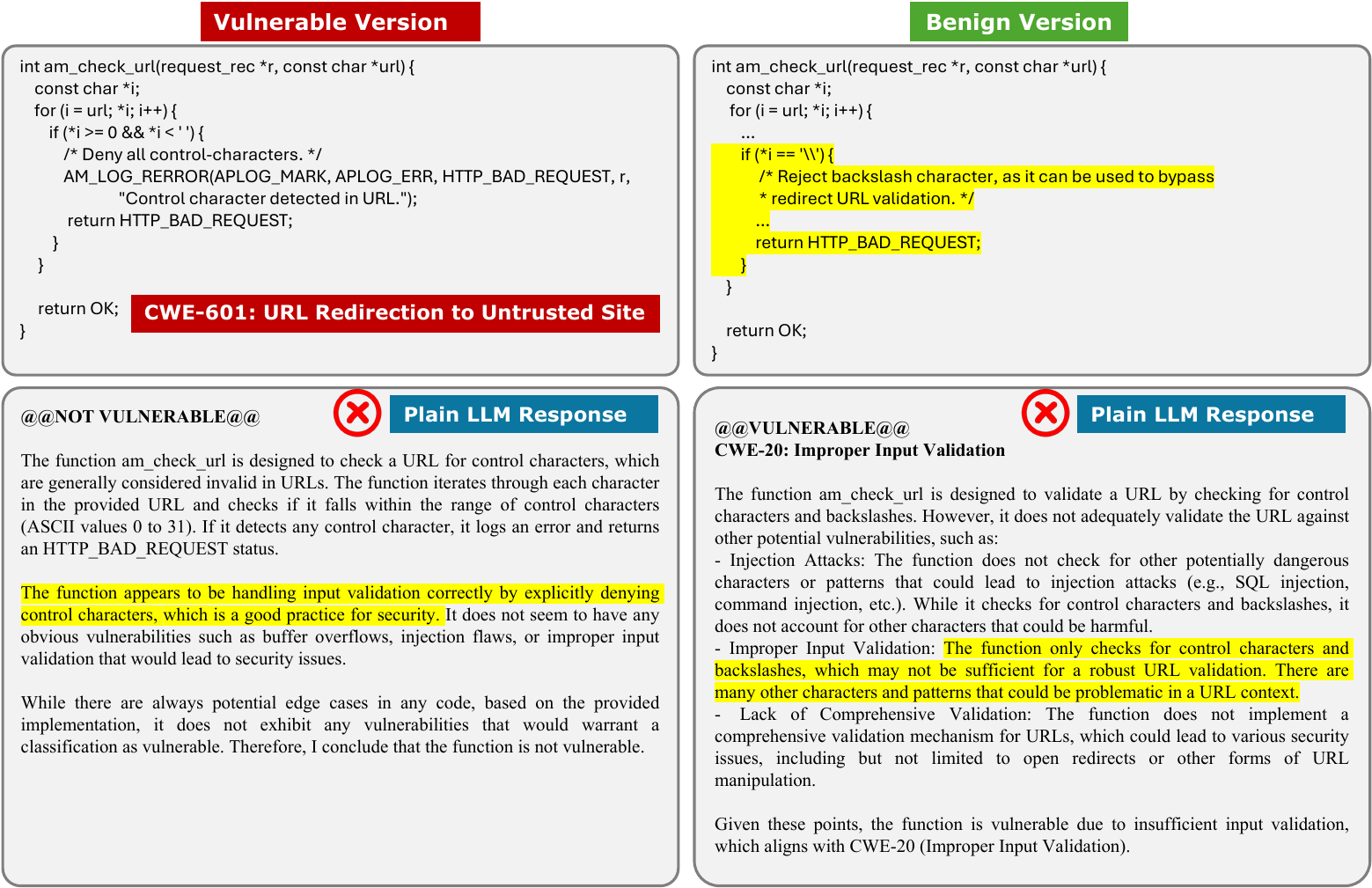}
    \caption{Case study for CVE-2019-3877.}\label{fig:case-2}
\end{figure}

Figure~\ref{fig:case-2} illustrates the case study derived from CVE-2019-3877 (details at https://nvd.nist.gov/vuln/detail/CVE-2019-3877), with the left side showing the vulnerable code and the detection methods' responses, and the right side depicting the benign version and its corresponding responses. The vulnerable version of the function \texttt{am\_check\_url} is susceptible to a ``CWE-601: URL Redirection to Untrusted Site'' vulnerability due to the insufficient validation for \texttt{url}. The benign version addresses this vulnerability by adding an if-condition to validate whether backslash exists in \texttt{url}.

\textbf{Label Predictions.} Plain LLM with GPT-4o and vanilla prompting produces incorrect predictions, often reversing the labels.

\textbf{Analysis Patterns.} A closer look at the analysis process reveals that Plain LLM fails to identify the insufficient validation for the \texttt{url} in the vulnerable version, where an assert is highlighted. In contrast, it over-analyzes the validation requirements for the benign version. These lead to incorrect predictions for both versions.

\end{document}